\documentclass[prb,twocolumn,showpacs]{revtex4}
\usepackage{graphicx}
\usepackage{amsmath}
\usepackage{amssymb}
\usepackage{dcolumn}
\usepackage{float}
\usepackage{bm}

\newcommand{\e}{\mathrm{e}}

\DeclareMathAlphabet{\bi}{OML}{cmm}{b}{it}

\def\be{\begin{equation}}
\def\ee{\end{equation}}
\def\bearr{\begin{eqnarray}}
\def\eearr{\end{eqnarray}}
\def\la{\langle}
\def\ra{\rangle}

\def\bs{\boldsymbol}

\begin{document}
\title{Phonon-drag thermopower and hot-electron energy-loss rate 
in a Rashba spin-orbit coupled two-dimensional electron system}
\bigskip

\author{Tutul Biswas and Tarun Kanti Ghosh}
\normalsize
\affiliation
{Department of Physics, Indian Institute of Technology-Kanpur,
Kanpur-208 016, India}
\date{\today}
 
\begin{abstract}
We theoretically study phonon-drag contribution to the 
thermoelectric power and hot-electron energy-loss rate  
in a Rashba spin-orbit coupled two-dimensional electron system 
in the Bloch-Gruneisen (BG) regime. 
We assume that electrons interact with longitudinal acoustic phonons 
through deformation potential and with both longitudinal and transverse
acoustic phonons through piezoelectric potential.
Effect of the Rashba spin-orbit interaction on magnitude and temperature 
dependence of the phonon-drag thermoelectric power and 
hot-electron energy-loss rate are discussed. 
We numerically extract the exponent 
of temperature dependence of the phonon-drag thermopower and the 
energy-loss rate. We find the exponents are suppressed due to 
the presence of the Rashba spin-orbit coupling.

\end{abstract}



\maketitle

\section{Introduction}

There has been a rapidly growing interest on spin-orbit 
coupled low-dimensional electron systems like 
two-dimensional electron systems (2DES) formed at semiconductor 
heterostructure interface, quantum wires, quantum dots 
\cite{zutic, winkler, cahay} etc.
The usefulness of the spin-orbit coupling in condensed matter system 
was realized after the proposal of spin field effect transistor 
by Datta and Das \cite{datta} and thereafter, various interesting 
theoretical and experimental studies have been reported.
One main goal is to control and manipulate the spin degree of
freedom of charge carriers in nanostructures so that 
spin-based electronic devices \cite{wolf,fabian,david} 
and quantum information processing technology can be developed. 
The coupling between electron's spin and orbital angular momentum
naturally arises when one makes non-relativistic approximation to the 
relativistic Dirac equation. 
There are various types of spin-orbit interaction (SOI) present in semiconductor
heterostructures. Most commonly used SOI is the 
Rashba SOI (RSOI) \cite{rashba}
which is due to the structural inversion asymmetry in semiconductor
heterostructures such as GaAs/AlGaAs.  
One can also increase the strength of the spin-orbit coupling 
with the help of an external gate voltage \cite{nitta,mats}.
The RSOI modifies various properties \cite{cape, grima, mars, 
chen, zhang} of a 2DES including the electron polaron effective mass,
polaron binding energy, mobility, etc. 

The electron-phonon interaction (EPI) plays an important
role in determining transport properties of a 2DES. 
There are mainly two kind of mechanisms responsible for the EPI in
semiconductor heterostructures, namely
deformation potential (DP) and piezoelectric (PE) scattering potential.
The deformation potential is the change in potential energy of an electron
due to lattice deformation.
In an inversion asymmetry host crystal, an electric polarization is
induced due to lattice vibration and the potential corresponds to 
the electric polarization is known as piezoelectric potential.
The electrons are scattered by the deformation and piezoelectric
potentials and provides a non-zero contribution to the
momentum relaxation time, in addition to other contributions coming from 
disorders, impurities etc.
Numerous studies have been devoted to probe EPI by measuring low-temperature
mobility \cite{price1, price2,price3,
price4, ridley, price5, dassarma} of a 2DES in the Bloch-Gruneisen (BG) regime.
The characteristic BG temperature\cite{stormer} is defined as $T_{BG}=2\hbar v_sk_F/k_B$
(where $v_s$ is the sound velocity and $k_F$ is the Fermi wave vector). When
the temperature decreases below $T_{BG}$ phonon modes with higher energy are
no longer able to be thermally excited and electrons are scattered by 
a small fraction of acoustic phonons with wave vector $q\leq2k_F$ due to the 
phase space restriction. This leads to a sharp decrease in resistivity 
$\rho\sim T^\nu$, where the exponent $\nu$ varies for different systems and 
different electron-phonon scattering mechanisms.
The BG temperature becomes $T_{BG}\sim6.2$ K for a typical electron density
$n_e\sim10^{11}$ m$^{-2}$ and sound velocity $v_s\sim5.12\times10^3$ ms$^{-1}$.

Thermoelectric properties of various materials including 2DES have 
attracted much interest due to potential applications.
With the application of an external temperature gradient 
${\bs \nabla}T$ across a sample, an electric field 
${\bf E} \propto {\bs \nabla}T$ is generated. 
The proportionality constant ($S$) between ${\bf E}$ and ${\bs \nabla}T$ 
is known as thermoelectric power or the Seebeck coefficient. 
There are mainly two contributions to the thermoelectric power $S$: 
diffusion thermopower $S_d$ and phonon-drag thermopower $S_g$. 
The applied temperature gradient gives rise to flow of electrons and
phonons from hotter region to cooler region. The diffusion thermopower
is solely due to flow of electrons and sensitive to the energy 
dependence of various scattering mechanisms such as ionized impurity
scattering, surface roughness scattering etc.
On the other hand, 
flow of phonons will try to drag the electrons from hotter region to 
cooler region due to EPI and giving 
rise to phonon-drag thermopower. 
Extensive theoretical and experimental investigations 
\cite{cantrell, cantrell1, cantrell2, lyo, karl, ruf, butcher, kuba, 
zianni, fletcher1, miele, butch, fletcher2, tsao, schm} on 
phonon-drag thermopower of a 2DES without RSOI have been performed.

Another relevant mechanism for probing the EPI 
is the energy-loss rate $(P$) of hot electrons. 
When an electron system is subjected to uniform heating, electron 
temperature raises above that of the phonons.
Hot electron relaxes to lower temperature via acoustic phonon emission. 
There are numerous experimental and theoretical studies 
\cite{dolgo, basu, sakaki,hess, kawamura, Ma, apple, pipa, stan,
kasa, pros} on hot-electron energy-loss 
rate of a 2DES without RSOI.

Unlike diffusion thermopower, both phonon-drag thermopower and 
hot-electron energy-loss rate depend only on EPI. 
Therefore, one can determine the electron-phonon 
coupling constant reliably by measuring $S_g$ and $P$.

Very recently, diffusive thermopower \cite{firoz} and acoustic 
phonon-limited resistivity \cite{ghosh} in a spin-orbit coupled 2DES 
have been studied.  
To the best of our knowledge, a detailed study of the effect of
RSOI on phonon-drag thermopower and hot-electron energy-loss rate 
have not been studied yet.
In this paper, we study phonon-drag contribution to the
thermoelectric power and energy-loss rate of a Rashba spin-orbit 
coupled quasi-2DES in which two-dimensional electron wave vector 
${\bf k}$ couples with three-dimensional phonon wave vector 
${\bf Q}=({\bf q},q_z)$. 
We consider both DP and PE scattering mechanisms responsible for 
the EPI. In the BG regime we find analytically 
that $S_g$ is proportional to $T^4$ and $T^2$ for DP and PE scattering,
respectively. On the other hand, approximate analytical calculations  
show that the energy-loss rate is proportional to $T^5$ and $T^3$
for DP and PE scattering, respectively. 
However, our numerical results reveal that the exponents 
are $strongly$ dependent on the electron density and the
Rashba spin-orbit coupling constant.

This paper is organized as follows. 
In section II we present all the analytical results
of phonon-drag thermopower and energy-loss rate. 
Numerical results and discussions have been reported 
in section III. We summarize our work in section IV.

\section{Theoretical details}
We consider a quasi-2DES formed at the interface of semiconductor 
heterostructures which has a finite thickness in the confining 
direction (say, $z$). Typically, the confining potential in the 
$z$-direction is a triangular potential. 
We assume that only the lowest sub-band due to transverse confinement 
is occupied by the electrons.
Therefore, electrons are restricted to move in the $xy$ plane
with wave vector ${\bf k}=(k_x,k_y)$. 
One can write the electron's wave function as
$\psi({\bf r})=\psi(x,y)\xi_0(z)$. 
The Fang-Howard wave function \cite{fang} in the $z$-direction is 
given by $\xi_0(z)=\sqrt{b^3/2}ze^{-bz/2}$ with 
$b=(48\pi m^\ast e^2/\varepsilon_0 \kappa\hbar^2)^{1/3}
\Big(n_d+11n_e/32\Big)^{1/3}$ 
as the variational parameter. 
Here, $ m^\ast $ is the effective mass of an electron,
$\kappa $ is the dielectric constant. Also, 
$\varepsilon_0$ is permittivity of free space, 
$n_d$ is the depletion charge density and 
$n_e$ is the density of electron.

The single electron Hamiltonian is given by
\begin{eqnarray}
H=\frac{{\bf p}^2}{2m^\ast} \sigma_0  +
\frac{\alpha}{\hbar}\big({\sigma}_x{p}_y-{\sigma}_y{p}_x\big),
\end{eqnarray}
where ${\bf p}=\hbar{\bf k}$ is the momentum operator for the electron,
$\sigma_0$ is the ${2\times 2}$ identity matrix, 
$\alpha$ is the RSOI coupling constant 
and $\sigma_{x(y)}$ are the usual Pauli spin matrices. 
The energy eigenvalues and the normalized eigenstates corresponding 
to the above Hamiltonian are, respectively, given by
\begin{equation} \label{energy}
\epsilon_k^{\lambda}=\frac{\hbar^2k^2}{2m^\ast} + 
\lambda \alpha k
\end{equation}
and
\begin{eqnarray} \label{spin states}
\psi_{\lambda} (x,y) = \frac{1}{\sqrt{2}}\begin{pmatrix} 1
\\ \lambda e^{-i \phi_k} \end{pmatrix} e^{i {\bf k} \cdot {\bf r}},
\end{eqnarray}
with $\lambda=\pm $ indicates two spin-split energy  branches 
and $\tan\phi_k=k_x/k_y$.
At a given Fermi energy $\epsilon_F$, the Fermi wave vectors for the 
two energy branches can be written as
$ k_F^\lambda = \sqrt{(k_F^0)^2 - k_{\alpha}^2}
-\lambda k_{\alpha} $
with $ k_F^0 = \sqrt{2\pi n_e} $ and 
$ k_{\alpha} = m^\ast\alpha/\hbar^2$.
The velocity of an electron in a particular branch $\lambda$ is given by
\begin{eqnarray}\label{velocity}
{v}_k^\lambda=\frac{1}{\hbar}\frac{\partial\epsilon_k^\lambda}{\partial k}
=\frac{\hbar k}{m^\ast}+\lambda\frac{\alpha}{\hbar}.
\end{eqnarray}

\subsection{Phonon-drag thermopower}
We consider the interaction between electrons with two-dimensional 
wave vector ${\bf k}$ and acoustic phonon with three-dimensional 
wave vector ${\bf Q}$.  
To calculate phonon-drag thermopower we follow the explicit formula 
given in References\cite{cantrell1, cantrell2} for 2DES. With appropriate modifications the 
expression for phonon-drag thermopower in a Rashba spin-orbit coupled 
2DES is given by
\begin{eqnarray} \label{phdrag1}
S_g^\lambda&=&\frac{e\tau_p}{2\sigma Ak_BT^2} 
\sum_{\lambda^\prime}\sum_{{\bf k}, 
{\bf k^\prime}, {\bf Q}} \hbar\omega_Q f(\epsilon_k^\lambda)
\Big[1-f(\epsilon_{k^\prime}^{\lambda^\prime})\Big]\nonumber\\
& \times & W_Q^{\lambda\lambda^\prime}({\bf k},{\bf k^\prime}) 
\Big\{\tau(\epsilon_k^\lambda){\bf v}_k^\lambda-
\tau(\epsilon_{k^\prime}^{\lambda^\prime})
{\bf v}_{k^\prime}^{\lambda^\prime}\Big\}\cdot{\bf v}_p,
\end{eqnarray}
where $e$ is the electronic charge, $\tau_p$ is the phonon 
mean free time, $A$ is the area of the sample, 
$\sigma$ is the Drude conductivity, $k_B$ is the Boltzmann constant, 
$\omega_Q=v_s Q$, $\tau(\epsilon_k)$ is the energy-dependent 
momentum relaxation time of an electron, 
$f(\epsilon) = [\e^{\beta(\epsilon-\mu)} + 1]^{-1}$ is the 
Fermi-Dirac distribution function with $\beta = 1/(k_{B}T)$, 
${\bf v}_k^\lambda$ is the velocity of an electron in a 
particular branch $\lambda$, ${\bf v}_p$ is the 
phonon velocity defined as ${\bf v}_p=v_s {\bf Q}/Q$ and
$W_Q^{\lambda\lambda^\prime}({\bf k},{\bf k^\prime})$ is the transition 
probability which is responsible for making transition of an electron 
from an initial state $\vert{\bf k},\lambda\ra$ to a final state 
$\vert{\bf k^\prime},\lambda^\prime\ra$ with the absorption of a phonon.
The explicit form of the transition probability is given by 
the Fermi's golden rule 
\begin{eqnarray}\label{trans_rate}
W_Q^{\lambda\lambda^\prime}({\bf k},{\bf k^\prime})=\frac{2\pi}{\hbar}
\vert C_ Q^{\lambda\lambda^\prime}\vert^2 N_{Q}
\delta\Big(\epsilon_{k^\prime}^{\lambda^\prime} - 
\epsilon_k^\lambda-\hbar\omega_Q\Big)
\delta_{{\bf k^\prime},{{\bf k}+{\bf q}}},
\end{eqnarray}
where $\vert C_ Q^{\lambda\lambda^\prime}\vert^2$ is the matrix 
element responsible for the EPI and  
$N_Q=[\exp(\beta \hbar\omega_Q)-1]^{-1}$ is the equilibrium phonon
distribution function.

The matrix elements for DP and PE
scatterings are respectively given by \cite{ghosh}  
\begin{equation}\label{dp_mat}
\vert C_{Q}^{\lambda,\lambda^{\prime}}\vert_{DP}^2=
\frac{D^2\hbar Q}{2\rho_mv_{sl}}
\frac{1+\lambda\lambda^{\prime}\cos\gamma_{{\bf k}{\bf k^\prime}}}{2} 
\delta_{\lambda,\lambda^\prime}
\vert I(q_z)\vert^2
\end{equation}
and
\begin{eqnarray} \label{piezo_mat}
\vert C_{Q,{l(t)}}^{\lambda,\lambda^{\prime}}\vert_{PE}^2 & = &
\frac{(eh_{14})^2\hbar}{2\rho_mv_{s{l(t)}}} 
\frac{1+\lambda\lambda^{\prime}\cos\gamma_{{\bf k}{\bf k^\prime}}}
{2\sqrt{q^2+q_z^2}} \delta_{\lambda,\lambda^\prime} \\ \nonumber 
& \times & \vert I(q_z)\vert^2A_{l{(t)}}({\bf q},q_z),
\end{eqnarray}
where $D$ is the DP coupling constant, $h_{14}$ is the 
relevant PE tensor component, $\rho_m$ is the mass density,
$v_{sl(t)}$ is the longitudinal (transverse) component of
sound velocity, $\gamma_{{\bf k}{\bf k^\prime}}$ is the angle
between ${\bf k}$ and ${\bf k^\prime}$,
$A_{l}({\bf q},q_z)=9q_z^2q^4/[2(q_z^2+q^2)^3]$ and 
$A_t({\bf q},q_z)=(8q_z^4q^2+q^6)/[4(q_z^2+q^2)^3]$. 
The Kronecker delta symbol $\delta_{\lambda,\lambda^\prime}$ in the
matrix elements implies that the EPI is
spin-independent.
Finally the 
form factor $\vert I(q_z)\vert^2$ which is responsible for the 
finite thickness of the quasi-2DES and it has the form
$\vert I(q_z)\vert^2=\vert\int dz\xi_0^2(z)e^{iq_zz}\vert^2 
=b^6/(q_z^2+b^2)^3$ for a triangular potential.

With the help of the Kronecker delta symbol 
$\delta_{{\bf k^\prime},{{\bf k}+{\bf q}}}$ in Eq. (\ref{trans_rate})
the summation over ${\bf k}^\prime$ in Eq. (\ref{phdrag1}) can be
easily evaluated by replacing ${\bf k^\prime}$ with ${{\bf k}+{\bf q}}$. 
The argument of the delta function in Eq. (\ref{trans_rate}) confirms 
the conservation of energy 
$\epsilon_{k^\prime}=\epsilon_{k}+\hbar\omega_Q$.
At this point we assume that $\tau(\epsilon_{k})$ is approximately 
constant over an energy scale of the order of $\hbar\omega_Q$ so that 
we can write $\tau(\epsilon_{k}+\hbar\omega_Q)\simeq\tau(\epsilon_{k})$.
The summation over ${\bf k}$ in Eq. (\ref{phdrag1}) can be converted into
an integral over $\epsilon_{k}$ by the following transformation
\begin{eqnarray}\label{trans}
\sum_{\bf k}\rightarrow\frac{A}{(2\pi)^2}\frac{m^\ast}{\hbar^2}
\int d\epsilon_{k}
\Big(1-\lambda\sqrt{\frac{\epsilon_\alpha}{\epsilon_{k}+\epsilon_\alpha}}
\Big)\int d\theta,
\end{eqnarray}
where $\theta$ is the angle between ${\bf k}$ and ${\bf q}$. 
At very low temperature we can make an additional approximation
as $f(\epsilon_{k})\{1-f(\epsilon_{k}+\hbar\omega_Q)\}\simeq
\hbar\omega_Q(N_Q+1)\delta(\epsilon_{k}-\epsilon_F)$.
Now we convert the summation over ${\bf Q}$ into an integration over
$q$ and $q_z$ as $\sum_{\bf Q}\rightarrow(1/4\pi^2)\int q dq dq_z$.
In BG regime, phonon energy is very small compared to the Fermi energy
and consequently we can make a further approximation as $q<<2k_F$.
With all these assumptions described above taken into account one can 
obtain a final expression for the phonon-drag thermopower as
(for intermediate steps see the Appendix A1)
\begin{eqnarray}\label{phndrag6}
S_g^\lambda&=&-\frac{e\tau_pv_sm^\ast\tau(\epsilon_F)}
{8\pi^3\hbar^4\sigma k_BT^2k_F^\lambda}
\Big(1-\lambda\sqrt{\frac{\epsilon_\alpha}{\epsilon_F+\epsilon_\alpha}}
\Big)\nonumber\\ 
&\times&\int dq dq_z q^2\frac{\vert C_Q^{\lambda,\lambda^\prime}\vert^2}{Q}
(\hbar\omega_Q)^2N_Q(N_Q+1).
\end{eqnarray}
Here, we have assumed $\tau^+(\epsilon_F)=\tau^-(\epsilon_F)=\tau(\epsilon_F)$
because the difference between $\tau^+$ and $\tau^-$ is very small.

The phonon energy is given by 
$\epsilon_p= \hbar \omega_Q = \hbar v_s\sqrt{q^2+q_z^2}$, so we can 
write $q=\epsilon_p\cos\phi/(\hbar v_s)$ and 
$q_z=\epsilon_p\sin\phi/(\hbar v_s)$ so that 
$dq dq_z\rightarrow \epsilon_p d\epsilon_p d\phi/(\hbar v_s)^2$.
We consider the quasi-2DES is very thin i.e. $q_z<<b$ so
$\vert I(q_z)\vert^2$ can be approximated as 
$\vert I(q_z)\vert^2 \simeq 1$.
With these substitutions $S_g$ due to DP scattering becomes
\begin{eqnarray}\label{phdrag6}
S_g^\lambda&=&-\frac{e\tau_pm^\ast\tau(\epsilon_F)}
{8\pi^3\hbar^4\sigma k_BT^2k_F^\lambda}
\Big(1-\lambda\sqrt{\frac{\epsilon_\alpha}{\epsilon_F+\epsilon_\alpha}}
\Big)\frac{D^2\hbar}{2\rho_m}\nonumber\\ 
& \times & 
\frac{1}{(\hbar v_{sl})^4}\int d\epsilon_p\epsilon_p^5N_Q(N_Q+1)
\int d\phi\cos^2\phi.
\end{eqnarray}
Using the standard result 
$\int d\epsilon_p\epsilon_p^nN_Q(N_Q+1)=(n)!\zeta(n)(k_BT)^{n+1}$
with $ \zeta(n)$  is the Riemann zeta function, 
we obtain 
\begin{eqnarray}
S_g^\lambda=-\frac{{m^\ast}^2 \Lambda D^2k_B}
{16\pi^2\rho_mn_e\hbar^2ek_F^\lambda}
\Big(1-\lambda\sqrt{\frac{\epsilon_\alpha}{\epsilon_F+\epsilon_\alpha}}
\Big)\frac{5!\zeta(5)(k_BT)^4}{(\hbar v_{sl})^5},
\end{eqnarray}
where $\Lambda=v_{sl}\tau_p$ is the phonon mean free path.
The total phonon-drag thermopower due to DP scattering is given by 
\begin{eqnarray}
S_g^{DP} = -\frac{{m^\ast}^2\Lambda D^2k_B}{8\pi^2\hbar^2\rho_mn_eek_F^0}
\sqrt{\frac{\epsilon_F^0}{\epsilon_F^0-\epsilon_\alpha}}
\frac{5!\zeta(5)(k_BT)^4}{(\hbar v_{sl})^5}.
\end{eqnarray}
The total phonon-drag thermopower $S_g$ for DP scattering
is proportional to $T^4$.

Calculations similar to DP scattering will yield 
the total phonon-drag thermopower for longitudinal
and transverse PE scatterings as

\begin{eqnarray}
S_{g,l}^{PE}=-\frac{45{m^\ast}^2\Lambda (eh_{14})^2k_B}
{2^{10}\pi^2\hbar^2\rho_mn_eek_F^0}
\sqrt{\frac{\epsilon_F^0}{\epsilon_F^0-\epsilon_\alpha}}
\frac{3!\zeta(3)(k_BT)^2}{(\hbar v_{sl})^3}
\end{eqnarray}
and 
\begin{eqnarray}
S_{g,t}^{PE}=-\frac{59 v_{st}{m^\ast}^2\Lambda 
(eh_{14})^2k_B}{2^{11} v_{sl}\pi^2\hbar^2\rho_mn_eek_F^0}
\sqrt{\frac{\epsilon_F^0}{\epsilon_F^0-\epsilon_\alpha}}
\frac{3!\zeta(3)(k_BT)^2}{(\hbar v_{st})^3}.
\end{eqnarray}

Total phonon-drag thermopower due to PE scattering 
is given by $S_g^{PE}=S_{g,l}^{PE}+2S_{g,t}^{PE}$ and 
in this case $S_g^{PE} \sim T^2$.

\subsection{Hot-electron Energy-loss rate}
 
The average energy-loss rate per electron via acoustic
phonon emission is given by
\begin{eqnarray} \label{enr_relx}
P=\Big\la\frac{\partial\epsilon}{\partial t}\Big\ra = 
\frac{1}{N_e} \sum_{\lambda, \bf Q} \hbar\omega_Q 
\Big[\frac{\partial N_Q}{\partial t}\Big]_\lambda,
\end{eqnarray}
where $N_e$ is the total number of electrons.
The rate of change of phonon occupation number for a given
branch $\lambda $ is given by
\begin{eqnarray}\label{phn_rate}
\Big[\frac{\partial N_Q}{\partial t}\Big]_\lambda&=&\sum_{\bf k}
W_Q^{\lambda\lambda}({\bf k}, {\bf k}+{\bf q})
\Big\{(N_Q+1)f(\epsilon_{k}^\lambda+\hbar\omega_Q)\nonumber\\
&\times&[1-f(\epsilon_{k}^\lambda)]
-N_Qf(\epsilon_{k}^\lambda)
[1-f(\epsilon_{k}^\lambda+\hbar\omega_Q)]\Big\},\nonumber\\  
\end{eqnarray}
where the Fermi-Dirac distribution function $f(\epsilon)$ is 
described by the temperature of hot electrons $T_e$ and phonon 
distribution function is described by the lattice temperature
$T_l$. Obviously $T_e$ is larger than $T_l$ so that electron 
can relax its energy via acoustic phonon emission 
and equilibrate to the lattice temperature.

Now we use the following identity
\begin{equation}\label{idn}
\frac{1-f(\epsilon_{\bf k}+\hbar\omega_Q)}{1-f(\epsilon_{\bf k})}
=\frac{f(\epsilon_{\bf k}+\hbar\omega_Q)}{f(\epsilon_{\bf k})}
e^{\beta_e\hbar\omega_Q},
\end{equation}
with $\beta_e=1/(k_BT_e)$.
Using Eqs. (\ref{enr_relx}) to (\ref{idn}) and taking 
all the assumptions made for calculating $S_g$ into 
account we finally obtain the following expression for 
energy-loss rate as (detail calculations are 
given in Appendix A2)
\begin{eqnarray}\label{energyloss2}
P&=&\frac{{m^\ast}^2}{4n_e\pi^3\hbar^5}
\sum_\lambda\frac{1}{k_F^\lambda}
\Big(1-\lambda\sqrt{\frac{\epsilon_\alpha}
{\epsilon_F+\epsilon_\alpha}} \Big)^2\int dq dq_z\nonumber\\
& \times & (\hbar\omega_Q)^2\vert C_Q^{\lambda,\lambda^{\prime}} \vert^2 
\Big\{\frac{1}{e^{\beta_e\hbar\omega_Q}-1}
-\frac{1}{e^{\beta_l\hbar\omega_Q}-1}\Big\},
\end{eqnarray}
with $\beta_l=1/(k_BT_l)$. The integration over $q$ and $q_z$ 
in Eq. (\ref{energyloss2}) can be easily evaluated by the same technique as 
described in the previous sub-section.
The expressions for total energy relaxation rate for 
DP, longitudinal PE and transverse PE scattering are 
respectively given by
\begin{eqnarray}\label{enlsDP}
P^{DP}=\frac{{m^\ast}^2D^2k_B^54!\zeta(5)}{2\pi^2n_e\rho_m\hbar^7v_{sl}^4k_F^0}
\sqrt{\frac{\epsilon_F^0}{\epsilon_F^0-\epsilon_\alpha}}(T_e^5-T_l^5),
\end{eqnarray}

\begin{eqnarray}
 P_l^{PE}=\frac{9{m^\ast}^2(eh_{14})^2k_B^32!\zeta(3)}
{64\pi^2n_e\rho_m\hbar^5v_{sl}^2k_F^0}
\sqrt{\frac{\epsilon_F^0}
{\epsilon_F^0-\epsilon_\alpha}}(T_e^3-T_l^3)
\end{eqnarray}
and
\begin{eqnarray}\label{enlsPEt}
P_{t}^{PE}=\frac{13{m^\ast}^2(eh_{14})^2k_B^32!\zeta(3)}
{128\pi^2n_e\rho_m\hbar^5v_{st}^2k_F^0}
\sqrt{\frac{\epsilon_F^0}{\epsilon_F^0-\epsilon_\alpha}}(T_e^3-T_l^3).
\end{eqnarray}

In the BG regime  $P^{DP}$ is proportional to $T^5$ for DP scattering 
and  $P^{PE} \sim T^3$ for PE scattering.

\section{Numerical Results}
In the previous section we have presented approximated analytical
results of the phonon-drag thermopower and energy-loss rate.
In this section we discuss numerical results of 
phonon-drag thermopower and hot-electron energy-loss rate. 
To do this we solve Eqs. (\ref{phdrag1}) and (\ref{enr_relx}) 
numerically for both DP and PE scattering mechanisms in the low
temperature regime. 
For the numerical calculations we consider
material parameters of GaAs/AlGaAs heterostructures as 
$m^\ast=0.067m_e$ with free electron mass $m_e$, 
$\rho_m=5.31\times10^3$ Kg m$^{-3}$,
$v_{sl}=5.12 \times 10^3$ ms$^{-1}$, 
$v_{st}=3.04\times10^3$ ms$^{-1}$,
$D=12$ eV, $h_{14}=1.2\times10^9$ Vm$^{-1}$, 
$\kappa=12.91$, $n_d=10^{14}$ m$^{-2}$,
$\alpha_0=10^{-11}$ eV m and $n_0=10^{15}$ m$^{-3}$.

\begin{figure}[ht]
\begin{center}\leavevmode
\includegraphics[width=110mm]{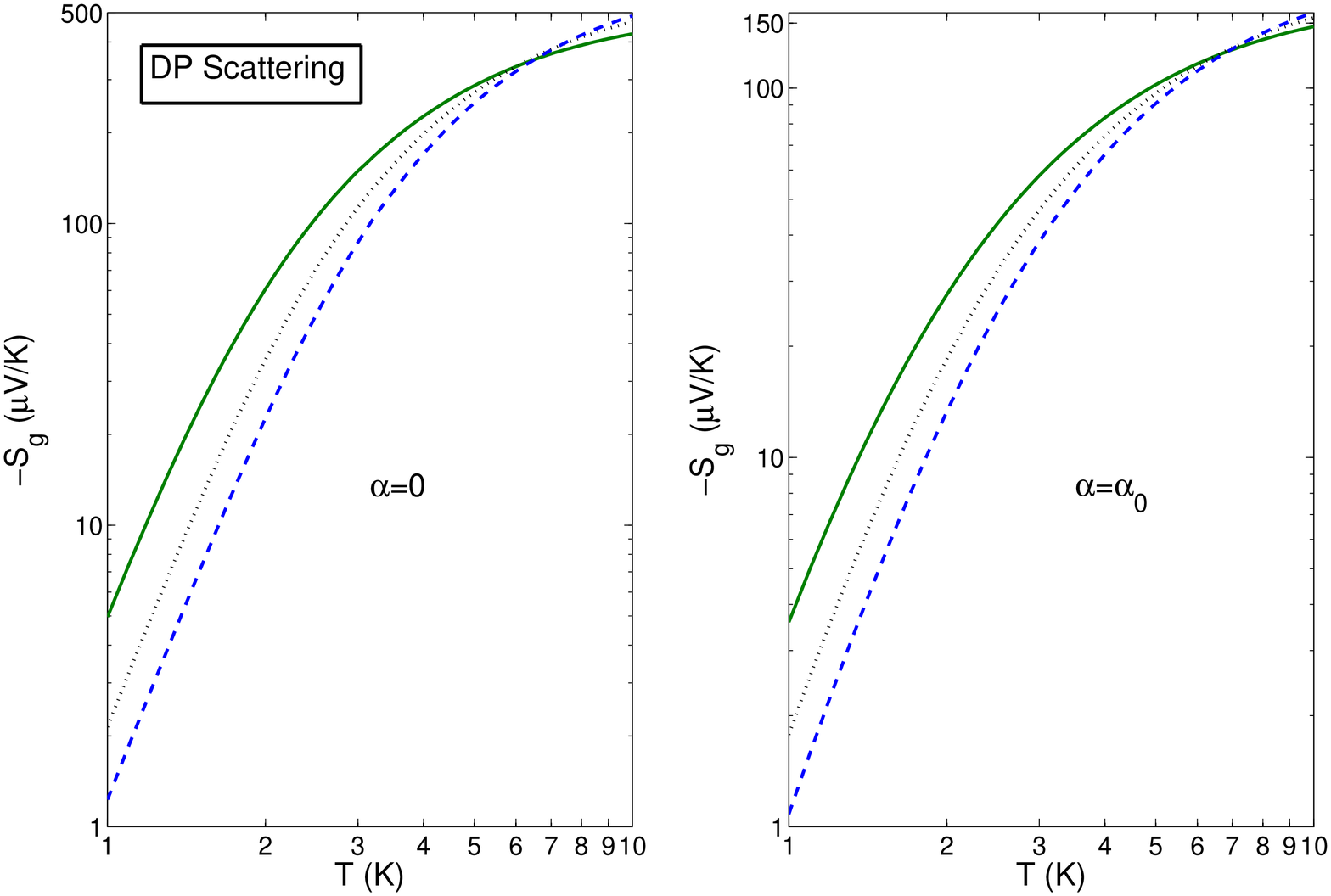}
\caption{(Color online) Plots of the phonon-drag thermopower due to 
DP scattering versus temperature for different values of the density.
Here, solid, dotted and dashed lines represent
$n_e = 3n_0$, $n_e=5n_0$ and $ n_e = 7n_0$, respectively.}
\label{Fig1}
\end{center}
\end{figure}

\begin{figure}[ht]
\begin{center}\leavevmode
\includegraphics[width=110mm]{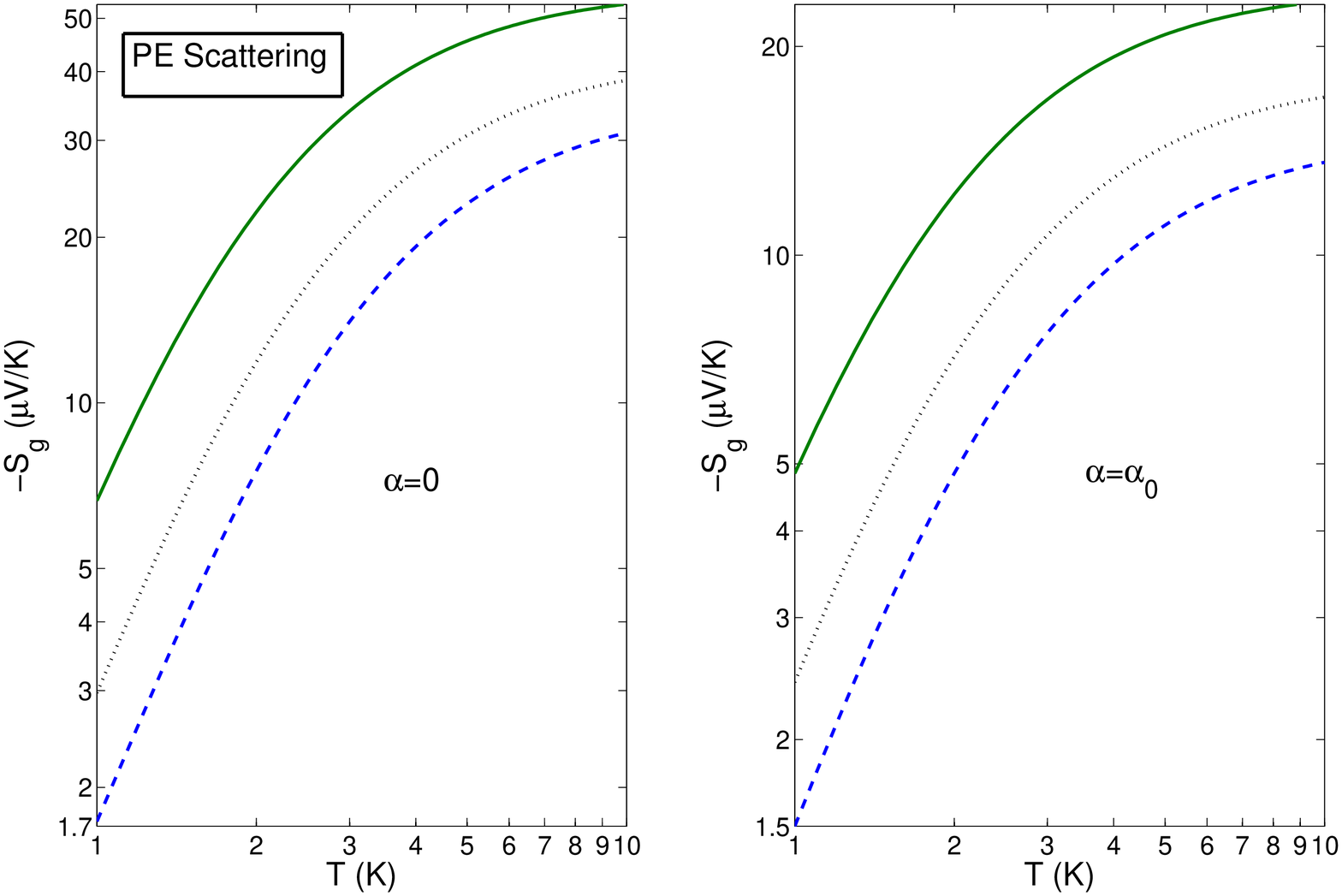}
\caption{(Color online) Plots of the phonon-drag thermopower due to PE 
scattering versus temperature for different values of the density.
Here, solid, dotted and dashed lines represent
$n_e = 3n_0$, $n_e=5n_0$ and $ n_e = 7n_0$, respectively.}
\label{Fig2}
\end{center}
\end{figure}

We estimate the effective exponent from the log-log plot of the 
phonon-drag thermopower versus temperature due to
DP scattering in Fig. 1  for both 
$\alpha=0$ and $\alpha=\alpha_0$ 
with different densities.
At very low temperature ($T \sim $ 1-3 K), we obtain 
$\nu=3.449, 3.942$ and $4.139 $ for $n_e=3n_0$, $5n_0$ and $7n_0$, 
respectively, when $\alpha =0$. 
On the other hand, we obtain 
$\nu=2.846, 3.294$ and $3.547$ for $n_e=3n_0$, $5n_0$ and $7n_0$, 
respectively, when $\alpha = \alpha_0$. 
The value of $\nu$ with $\alpha=\alpha_0$ gets lowered than that 
with $\alpha=0$. 
Figure 1 also depicts that the magnitude of $S_g$ with 
$\alpha=\alpha_0$ is less in comparison with $\alpha=0$.

In Fig. 2, the log-log plot of the phonon-drag thermopower versus 
temperature due to PE scattering is presented for different
values of $\alpha $ and $n_e$. 
When $\alpha =0$, we obtain $\nu=1.688, 1.961$ and $2.096$ for $n_e=3n_0$,
$5n_0$ and $7n_0$, respectively. When $\alpha=\alpha_0$, we obtain
$\nu=1.290, 1.520$ and $1.658$ for $n_e=3n_0$, $5n_0$ and $7n_0$, 
respectively.
Comparing Fig. 1 and Fig. 2, one can conclude that the magnitude of 
$S_g$ due to DP scattering is larger than that due to PE scattering.

\begin{figure}[ht]
\begin{center}\leavevmode
\includegraphics[width=110mm]{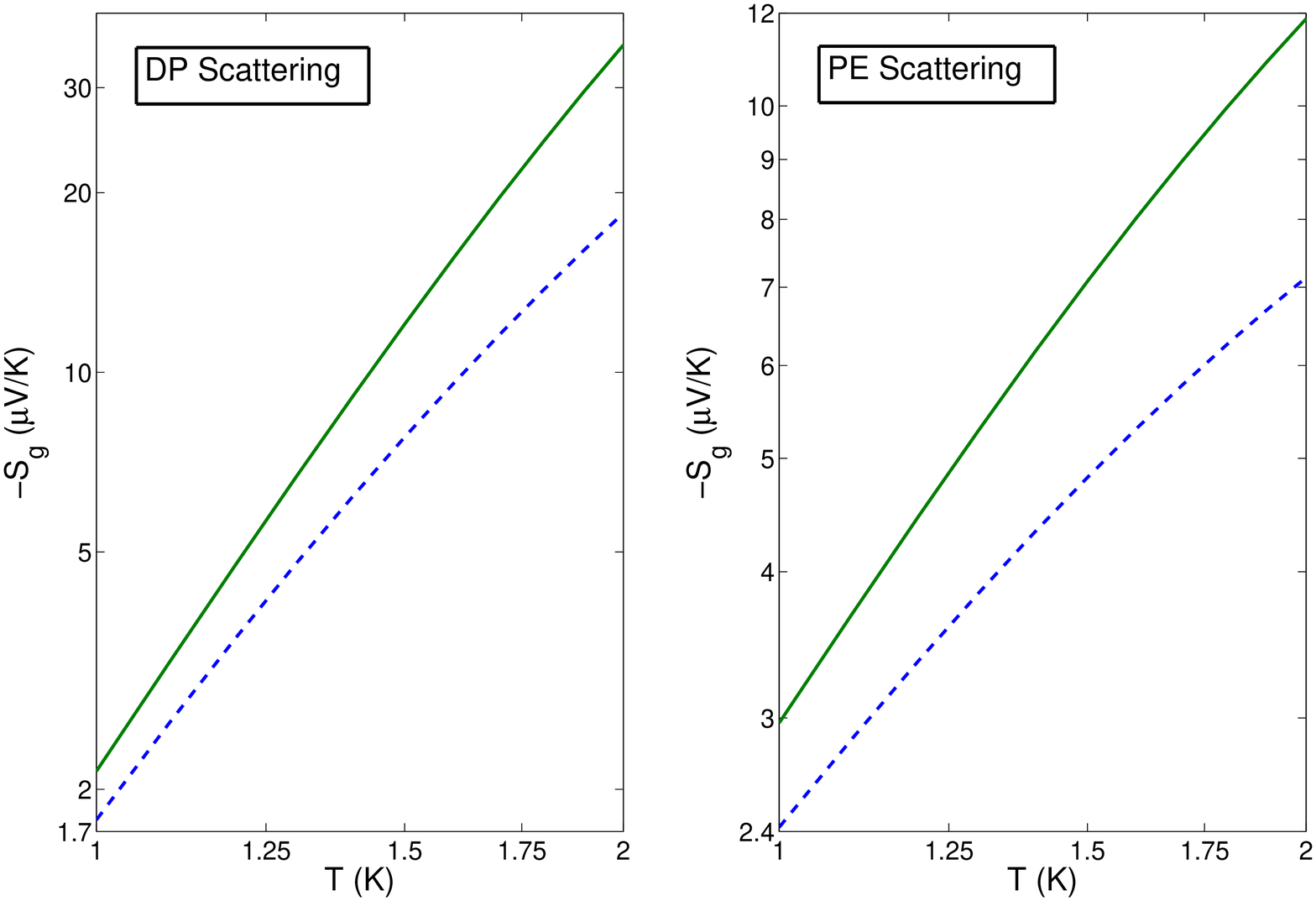}
\caption{(Color online) Plots of the phonon-drag thermopower due to DP 
and PE scattering versus temperature for a fixed density $n_e = 5n_0$.
Here, solid and dashed lines represent $\alpha =0 $ and 
$\alpha = \alpha_0$, respectively.}
\label{Fig3}
\end{center}
\end{figure}

In Fig. 3 we plot $S_g$ due to DP and PE scattering versus $T$
for $\alpha=0$ and $\alpha=\alpha_0$ at a fixed density $n_e=5n_0$.
Figure 3 clearly shows that slope of the line with
$\alpha=\alpha_0$ is less than that with $\alpha=0$ case.

\begin{figure}[ht]
\begin{center}\leavevmode
\includegraphics[width=110mm]{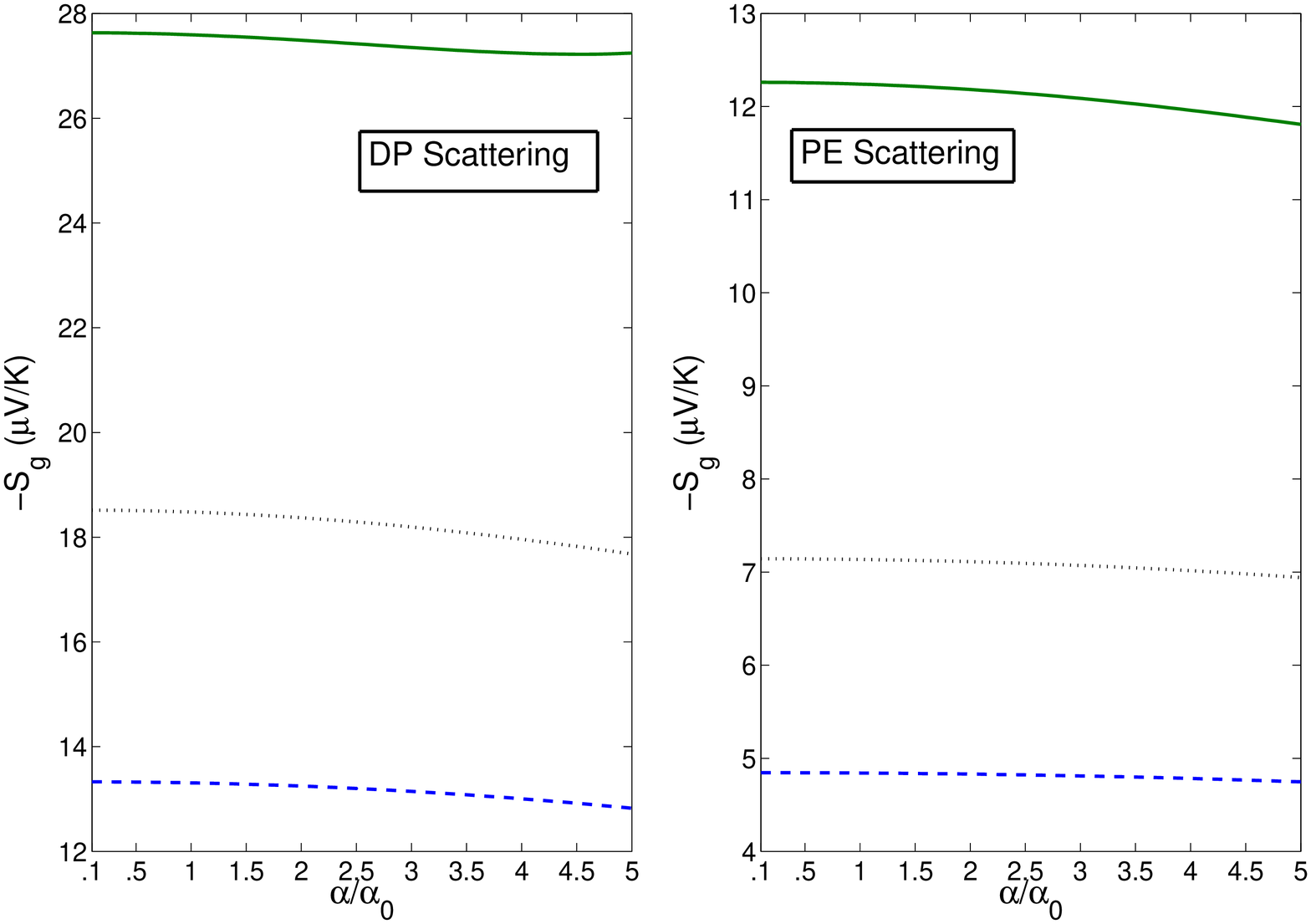}
\caption{(Color online) Plots of the phonon-drag thermopower due to DP 
and PE scattering versus $\alpha$ for different values of the density
at fixed temperature $T=2$K.
Here, solid, dotted and dashed lines represent
$n_e = 3n_0$, $n_e=5n_0$ and $ n_e = 7n_0$, respectively.}
\label{Fig4}
\end{center}
\end{figure}

\begin{figure}[ht]
\begin{center}\leavevmode
\includegraphics[width=110mm]{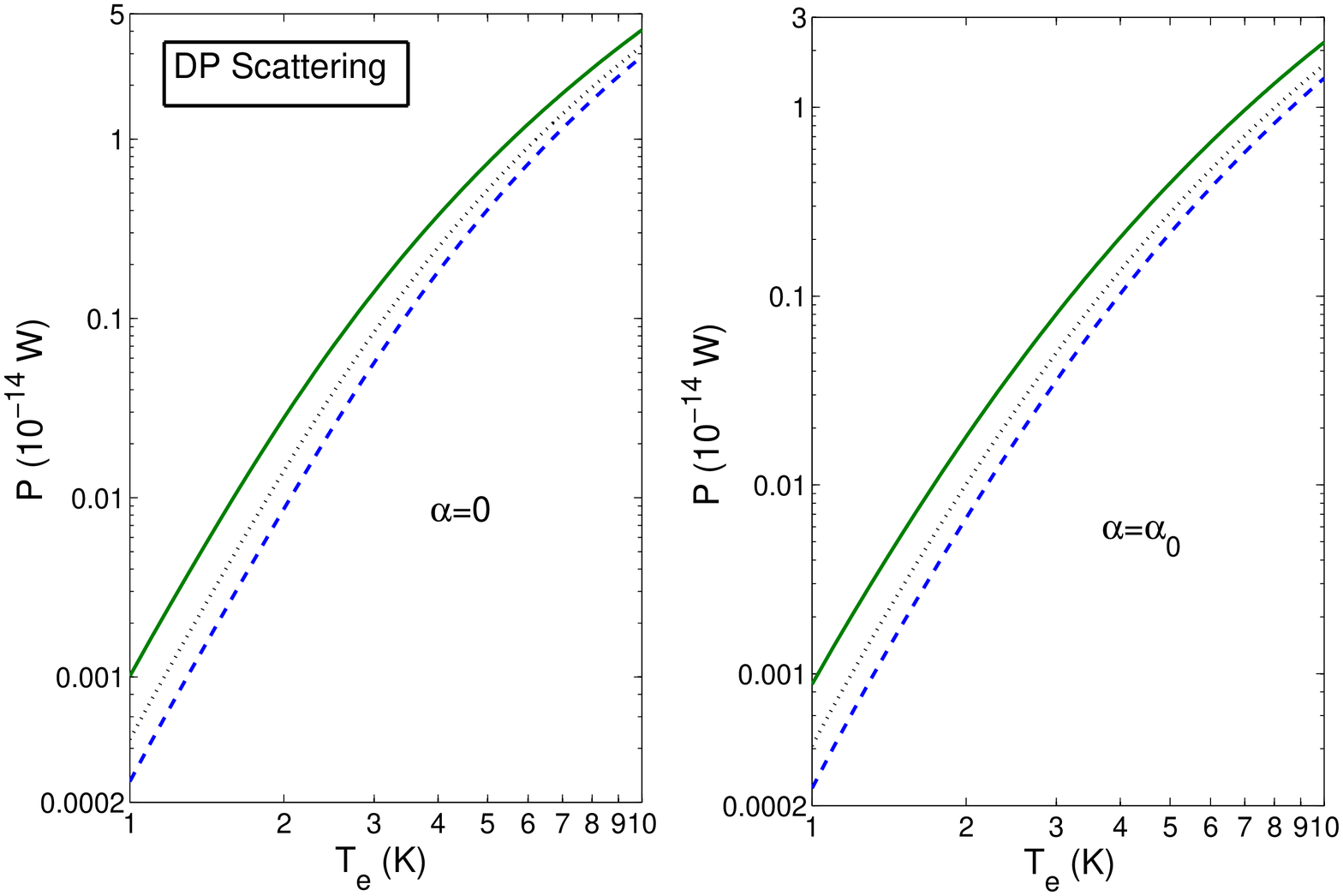}
\caption{(Color online) Plots of the energy-loss rate due to 
DP scattering as a function of electron temperature $T_e$ for 
$\alpha=0$ and $\alpha=\alpha_0$ for different density. 
We set the lattice temperature $T_l=0$.
Here, solid, dotted and dashed lines represent
$n_e = 3n_0$, $n_e=5n_0$ and $ n_e = 7n_0$, respectively.}
\label{Fig5}
\end{center}
\end{figure}

Variation of $S_g$ due to DP and PE scattering as a function
of the the Rashba coupling constant ($\alpha$) is shown in Fig. 4.
The phonon-drag thermopower $S_g$ decreases very slowly with the increase of 
$\alpha$ in the case of both DP and PE scattering.

It is important to compare the order of magnitudes of 
diffusion and phonon-drag contributions to the total 
thermopower. The diffusion thermopower \cite{firoz} is given by 
\begin{eqnarray}\label{diffusion}
S_d=-\frac{\pi^2k_B^2T}{3\vert e\vert\epsilon_F}\Big(p+1
-\frac{\epsilon_\alpha}{\epsilon_F}\Big),
\end{eqnarray}
where the parameter $p$ depends on various scattering mechanisms.
We calculate $S_d$ from Eq. (\ref{diffusion})
and $S_g$ from numerical evaluation of Eq. (\ref{phdrag1}).
With $\alpha=\alpha_0$ and $n=5n_0$ at $T=2$ K we obtain 
$S_d\sim -5.464 $ $\mu$V$/$K, 
$S_g^{DP} \sim -18.526 $ $\mu$V$/$K and 
$S_g^{PE} \sim -7.135 $ $\mu$V$/$K. 
It is clear that at $T = 2$ K 
the phonon-drag thermopower due to DP scattering 
dominates over both PE scattering induced 
phonon-drag thermopower and diffusion thermopower.

\begin{figure}[ht]
\begin{center}\leavevmode
\includegraphics[width=110mm]{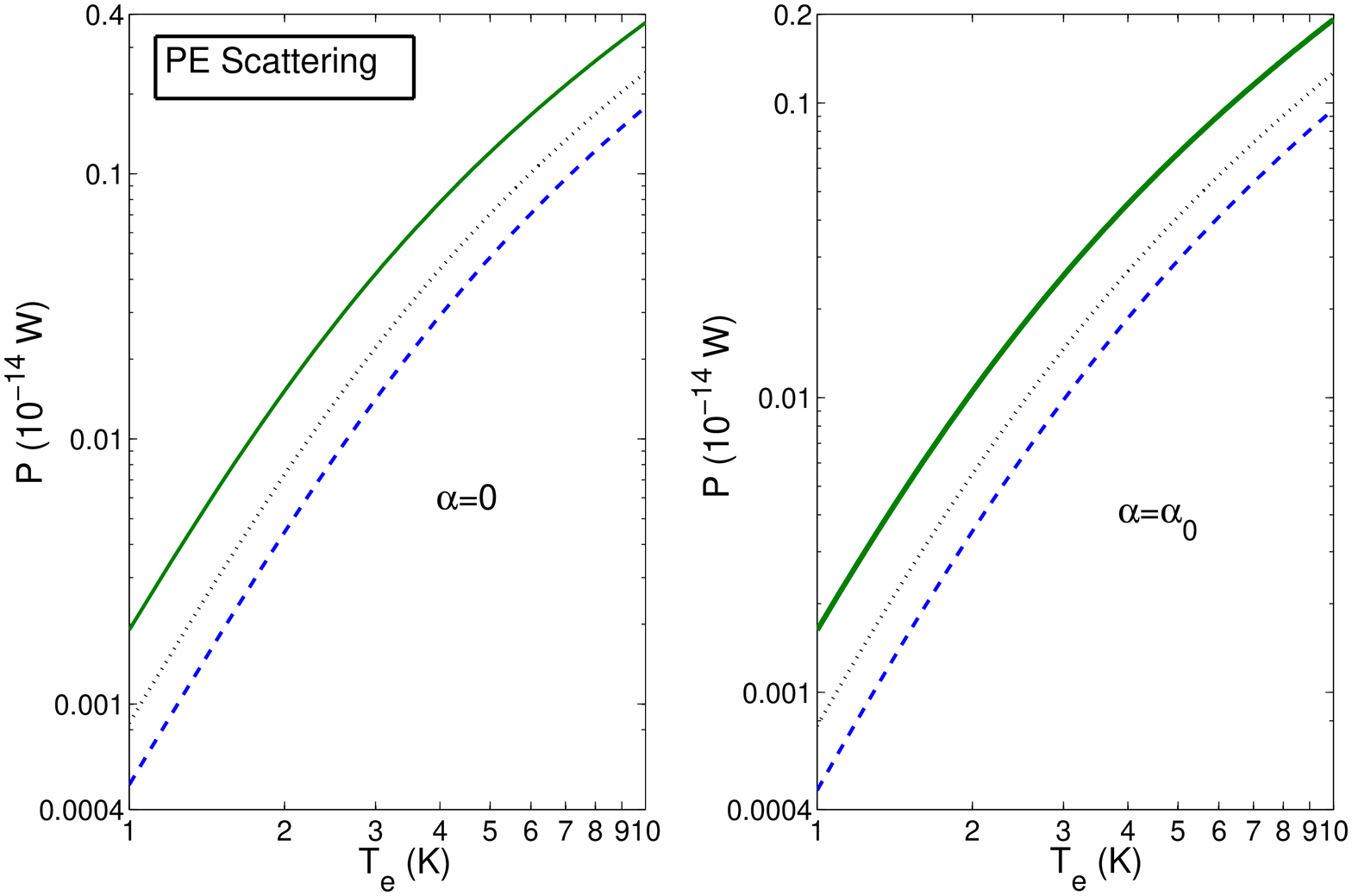}
\caption{(Color online) Plots of the energy-loss rate due to PE 
scattering versus electron temperature $T_e$ for 
$\alpha=0$ and $\alpha=\alpha_0$ for different density. 
Here, lattice temperature is fixed to $T_l=0$.
Solid, dotted and dashed lines represent
$n_e = 3n_0$, $n_e=5n_0$ and $ n_e = 7n_0$, respectively.}
\label{Fig7}
\end{center}
\end{figure}

\begin{figure}[ht]
\begin{center}\leavevmode
\includegraphics[width=110mm]{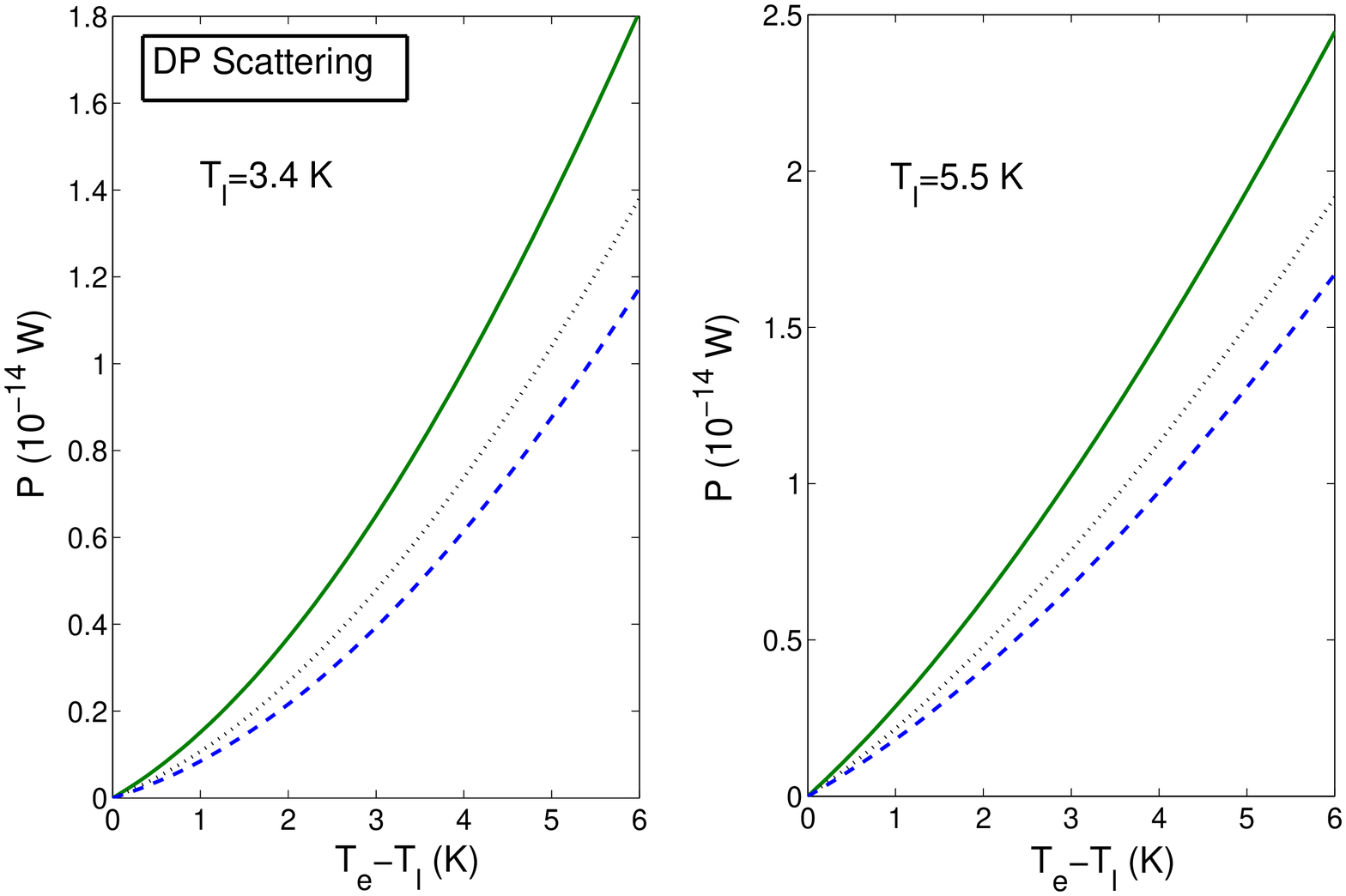}
\caption{(Color online) Energy-loss rate due to DP scattering
is plotted as a function of $T_e-T_l$ with $\alpha=\alpha_0$ for two
lattice temperatures $T_l=3.4$ K and $T_l=5.5$ K.
Here, solid, dotted and dashed lines represent
$n_e = 3n_0$, $n_e=5n_0$ and $ n_e = 7n_0$, respectively.}
\label{Fig7}
\end{center}
\end{figure}

\begin{figure}[ht]
\begin{center}\leavevmode
\includegraphics[width=110mm]{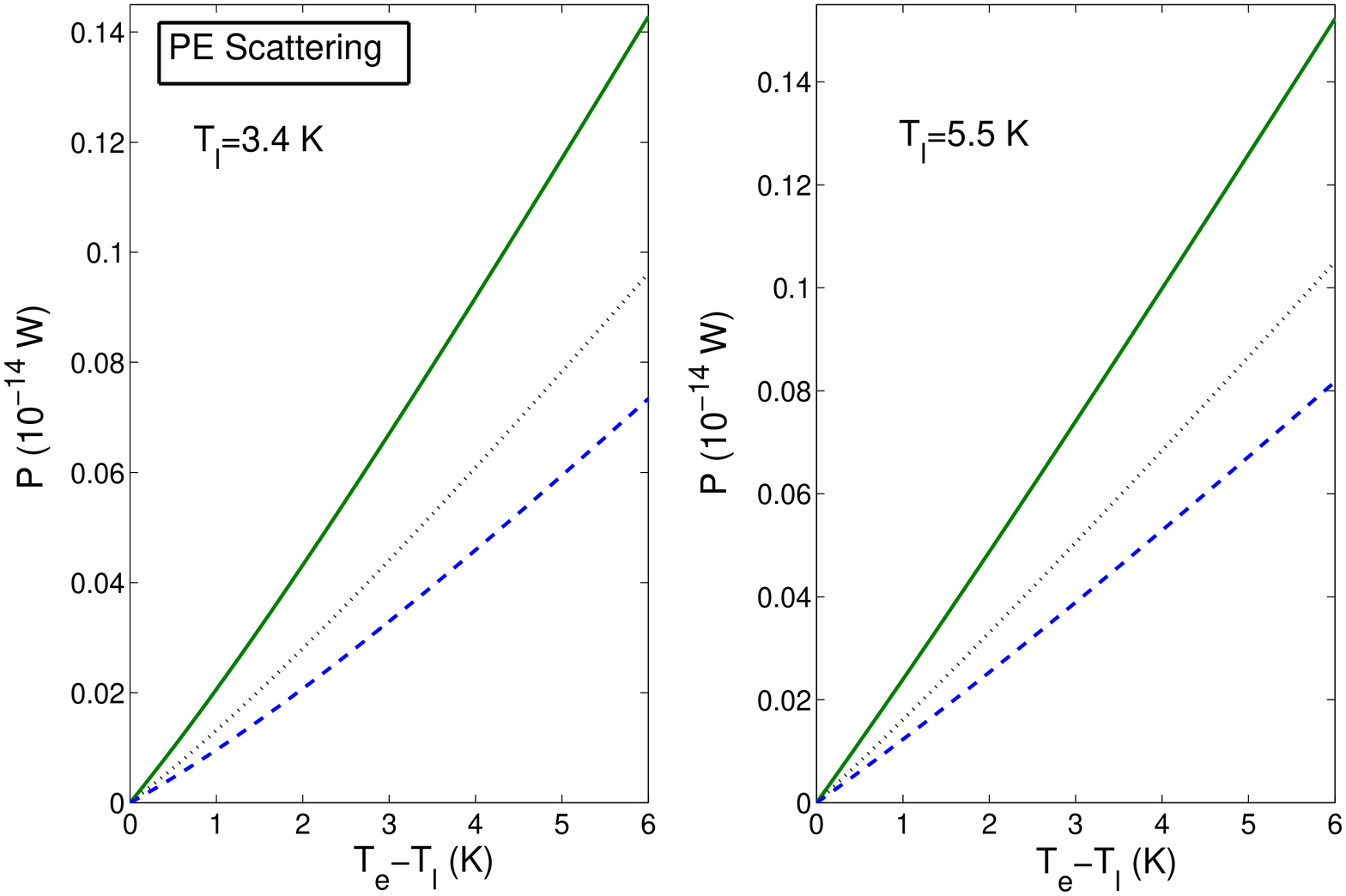}
\caption{(Color online) Energy-loss rate due to PE scattering
is plotted as a function of $T_e-T_l$ with $\alpha=\alpha_0$ for two
lattice temperatures $T_l=3.4$ K and $T_l=5.5$ K.
Here, solid, dotted and dashed lines represent
$n_e = 3n_0$, $n_e=5n_0$ and $ n_e = 7n_0$, respectively.}
\label{Fig7}
\end{center}
\end{figure}

\begin{figure}[ht]
\begin{center}\leavevmode
\includegraphics[width=110mm]{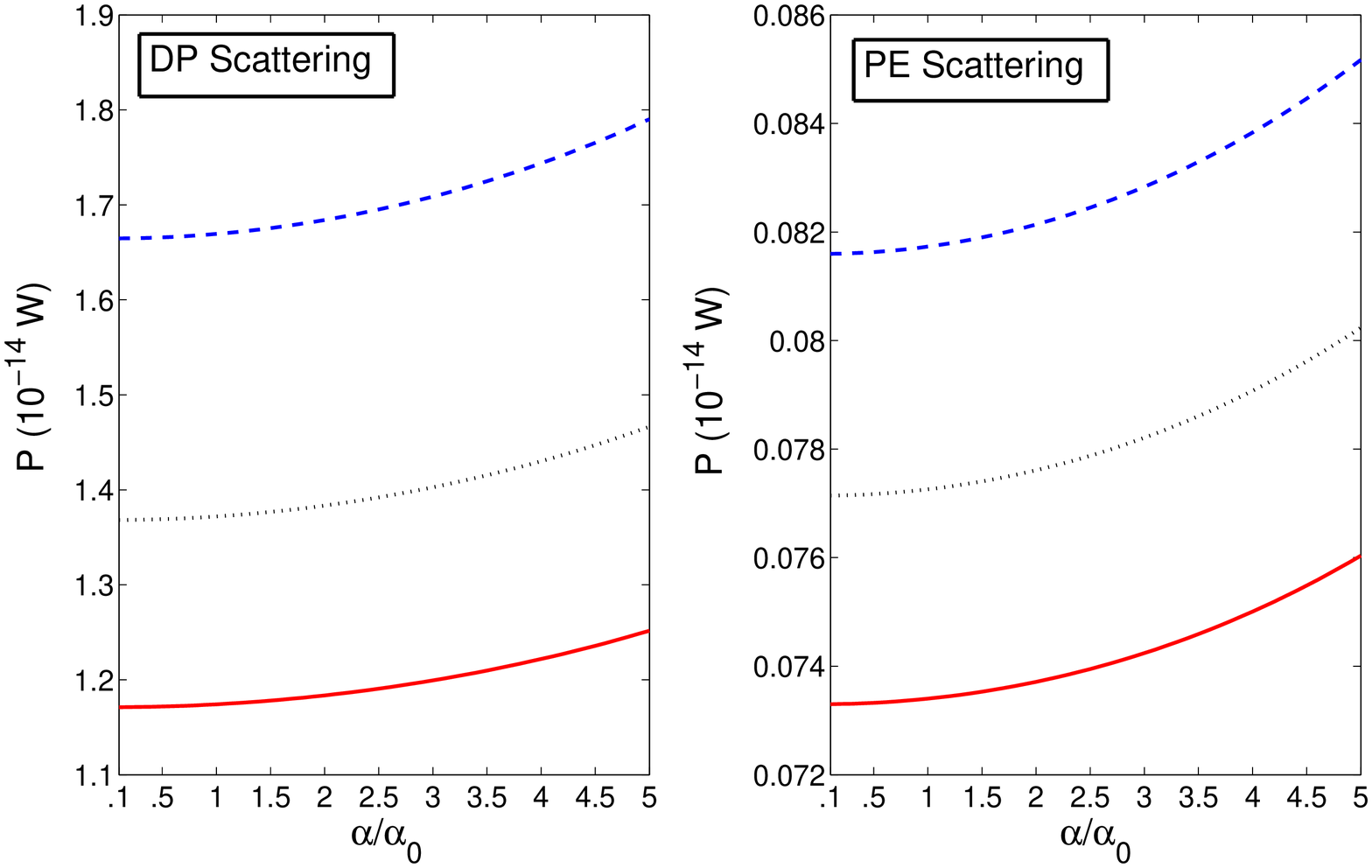}
\caption{(Color online) Plots of the energy-loss rate due to DP and PE 
scattering
versus $\alpha$ for different values of the lattice temperature $T_l$.
We fix the electron temperature at $T_e=6$ K.
Here, solid, dotted and dashed lines represent
$T_l=3.4$ K, $T_l=4.2$ K and $T_l=5.5$ K, respectively.}
\label{Fig7}
\end{center}
\end{figure}

Now we turn to present the numerical calculations for 
energy-loss rate of hot electrons in the BG regime.
In Figs. 5 and 6 we have shown the dependence of 
$P$ with electron temperature $T_e$ for DP and 
PE scattering, respectively. 

In general, energy-loss rate of hot-electrons is given by
$P=\Gamma(T_e^\delta-T_l^\delta)$ with $\Gamma$ as the proportionality
constant. By taking the lattice temperature $T_l=0$ we  
determine the values of $\delta$ for $\alpha=0$ and
$\alpha=\alpha_0$ from Figs. 5 and 6. For DP scattering with
$\alpha=0$ the values of $\delta$ are $\delta=4.707, 4.916$ and $4.998$  
for $n_e=3n_0, 5n_0$ and $7n_0$, respectively. When $\alpha=\alpha_0$ we 
obtain $\delta=4.268, 4.511$ and $4.661$  
for $n_e=3n_0, 5n_0$ and $7n_0$, respectively. 
For PE scattering we find $\delta=2.945, 3.078$ and $3.139$  
for $n_e=3n_0, 5n_0$ and $7n_0$, respectively at $\alpha=0$.
With $\alpha=\alpha_0$ the values of $\delta$ are obtained as 
$\delta=2.635, 2.790$ and $2.880$  
for $n_e=3n_0, 5n_0$ and $7n_0$, respectively.

Our numerical calculations reveal that the effective 
exponents of temperature dependence of phonon-drag 
thermopower and energy-loss rate strongly depend on the
electron density and the Rashba spin-orbit coupling constant.

In Figs. 7 and 8 we plot $P$ due to DP and PE scattering
as a function of $T_e-T_l$ with $\alpha=\alpha_0$ for
different density. Different values of $T_l$ have been 
considered here. Magnitude of $P$ is higher at higher 
values of $T_l$ and it decreases with increase of the electron density.
Comparing Figs. 7 and 8, we see that the energy-loss rate due
to DP scattering is much higher than that due to PE case.

In Fig. 9, we present energy-loss rate due to DP and PE scattering
versus $\alpha$ for different lattice temperatures. It shows 
the energy-loss rates increase monotonically with $\alpha$ in both the
cases. This behavior is quite different from the 
behavior of the phonon-drag thermopower versus $\alpha$ shown in Fig. 4.

\begin{table}[ht]
\centering
\begin{tabular}{|c|c|c|c|c|}\hline

\multicolumn{1}{|c|}{Density} &
\multicolumn{2}{c|}{DP} &
\multicolumn{2}{c|}{PE} \\ \cline{2-5}

\multicolumn{1}{|c|}{$(n_e)$} &
\multicolumn{1}{|c|}{$\alpha=0$} &
\multicolumn{1}{|c|}{$\alpha=\alpha_0$} &
\multicolumn{1}{|c|}{$\alpha=0$} &
\multicolumn{1}{|c|}{$\alpha=\alpha_0$} \\ \hline

\ \ \ 3$n_0$ & 3.449 & 2.846 & 1.688 & 1.290 \\
\ \ \ 5$n_0$ & 3.942 & 3.294 & 1.961 & 1.520 \\
\ \ \ 7$n_0$ & 4.139 & 3.547 & 2.096 & 1.658 \\ \hline

\end{tabular}
\caption{The effective exponent of the temperature dependence of 
$S_g$ in the BG regime for various values of $n_e$ and $\alpha$.}
\end{table}

\begin{table}[h!]
\centering
\begin{tabular}{|c|c|c|c|c|}\hline

\multicolumn{1}{|c|}{Density} &
\multicolumn{2}{c|}{DP} &
\multicolumn{2}{c|}{PE} \\ \cline{2-5}

\multicolumn{1}{|c|}{$(n_e)$} &
\multicolumn{1}{|c|}{$\alpha=0$} &
\multicolumn{1}{|c|}{$\alpha=\alpha_0$} &
\multicolumn{1}{|c|}{$\alpha=0$} &
\multicolumn{1}{|c|}{$\alpha=\alpha_0$} \\ \hline

\ \ \ 3$n_0$ & 4.707 & 4.268 & 2.945 & 2.635 \\
\ \ \ 5$n_0$ & 4.916 & 4.511 & 3.078 & 2.790 \\
\ \ \ 7$n_0$ & 4.998 & 4.661 & 3.139 & 2.880 \\ \hline

\end{tabular}
\caption{The effective exponent of the temperature dependence of 
$P$ in the BG regime for various values of $n_e$ and $\alpha$.}
\end{table}

\section{Summary}

In this section we are summarizing the main results of 
the present work. In BG regime phonon-drag thermopower and hot-electron 
energy-loss rate have been calculated for quasi-2DES formed
at the interface of GaAs/AlGaAs heterojunction. Both DP and PE
scattering mechanism have been taken into account separately.
It is shown that the effective exponent of the temperature dependence
of the phonon-drag thermopower and energy-loss rate strongly
depend on the electron density and the Rashba spin-orbit coupling 
constant. For DP and PE scattering in the BG regime, the values of 
the effective exponents of $T$ of thermoelectric power and 
hot-electron energy-loss rate for different values of $n_e$ 
and $\alpha$ are summarized in Table I and Table II.

It is shown that the order of magnitudes
of phonon-drag thermopower due to DP and PE scattering are almost same 
and the phonon-drag thermopower dominates over
the diffusion thermopower in the BG regime. 
The phonon-drag thermopower due to DP and PE scattering decreases very slowly with
increase of $\alpha$.

The order of magnitude of the energy-loss rate due to DP scattering
is much higher than that of the PE scattering. The energy-loss rate due 
to DP and PE scattering increases monotonically with $\alpha$.

\appendix

\section{}
In this appendix, we briefly sketch the derivation of the phonon-drag 
thermopower.
Using Eqs. (\ref{phdrag1}) and (\ref{trans}) we can write
\begin{eqnarray}\label{phdrag3}
S_g^\lambda&=&\frac{e\tau_pm^\ast}{8\pi^2\hbar^2\sigma k_BT^2}
\sum_{\bf Q} \int d\epsilon_{k} 
\Big(1-\lambda\sqrt{\frac{\epsilon_\alpha}
{\epsilon_{k}+\epsilon_\alpha}}\Big)\nonumber\\
& \times & \int d\theta\hbar\omega_Q 
W_Q^{\lambda\lambda}({\bf k},{{\bf k}+{\bf q}}) f(\epsilon_{k})
\Big\{1-f(\epsilon_{k}+\hbar\omega_Q)\Big\}\nonumber\\
&\times&\tau(\epsilon_{k}^\lambda)\Big({\bf v}_{k}^\lambda
-{\bf v}_{{\bf k}+{\bf q}}^\lambda\Big)\cdot{\bf v}_p.
\end{eqnarray}

Using the approximation 
$f(\epsilon_{k})\{1-f(\epsilon_{k}+\hbar\omega_Q)\} 
\simeq
\hbar\omega_Q(N_Q+1)\delta(\epsilon_{k}-\epsilon_F)$,
the integration over $\epsilon_{k}$ in  Eq. (\ref{phdrag3}) 
can be easily done and we obtain
\begin{eqnarray}\label{phdrag4}
S_g^\lambda&=&\frac{e\tau_pm^\ast \tau(\epsilon_F)}
{8\pi^2\hbar^2\sigma k_BT^2}
\Big(1-\lambda\sqrt{\frac{\epsilon_\alpha}{\epsilon_F+\epsilon_\alpha}}
\Big)\sum_{{\bf Q}}\int d\theta(\hbar\omega_Q)^2\nonumber\\ 
&\times&(N_Q+1)
W_Q^{\lambda\lambda}({\bf k}_F,{{\bf k}_F+{\bf q}})\Big({\bf v}_{{\bf k}_F}^\lambda
-{\bf v}_{{{\bf k}_F}+{\bf q}}^\lambda\Big)\cdot{\bf v}_p.\nonumber\\
\end{eqnarray}

Using Eq. (\ref{velocity}) and taking $q<<k_F$ we can write 
the difference between two velocities as
\begin{eqnarray}
{\bf v}_{{\bf k}_F}^\lambda-{\bf v}_{{{\bf k}_F}+{\bf q}}^{\lambda}
\simeq -\frac{\hbar{\bf q}}{m^\ast}-\lambda\frac{\alpha}{\hbar}
\Big(\frac{{\bf q}}{k_F}-\frac{{{\bf k}_F}({\bf k}_F\cdot{\bf q})}{k_F^3}\Big).
\end{eqnarray}

Similarly, the difference between two energies can be approximated as
\begin{eqnarray}
\epsilon_{{\bf k}_F + {\bf q}}^{\lambda} - 
\epsilon_{{\bf k}_F}^\lambda
& \simeq & 
\Big(\frac{\hbar^2k_F}{m^\ast} + \lambda \alpha \Big) 
q\cos\theta\nonumber \\ 
&+& 
\Big(\frac{\hbar^2}{2m^\ast}+ 
\lambda\frac{\alpha}{2k_F}\Big)q^2,
\end{eqnarray}
where $\theta$ is the angle between ${\bf k}_F$ and ${\bf q}$.
Converting the summation over ${\bf Q}$ into an integration over 
$q$, $q_z$ and doing the $\theta$ integration we can simplify 
Eq. (\ref{phdrag4}) as
\begin{eqnarray}\label{phdrag5}
S_g^\lambda&=&-\frac{e\tau_pv_s{m^\ast}^2\tau(\epsilon_F)}
{8\pi^3\hbar^5\sigma k_BT^2{k_F^\lambda}^2}
\Big(1-\lambda\sqrt{\frac{\epsilon_\alpha}{\epsilon_F+\epsilon_\alpha}}
\Big)^2\int dq dq_z q^2 \nonumber\\ 
&\times&(\hbar\omega_Q)^2N_Q(N_Q+1)\frac{\vert C_Q\vert^2}{QG(k_F^\lambda,Q)}
\Big[\frac{\hbar k_F^\lambda}{m^\ast}+\lambda\frac{\alpha}{\hbar}\nonumber\\
&-&\lambda\frac{\alpha}{\hbar}
\Big\{\frac{m^\ast v_s}{\hbar k_F^\lambda}\frac{Q}{q}
\Big(1-\lambda\sqrt{\frac{\epsilon_\alpha}{\epsilon_F+\epsilon_\alpha}}
\Big)-\frac{q}{2k_F^\lambda}\Big\}^2\Big],
\end{eqnarray}
with
\begin{eqnarray}
G(k_F^\lambda,Q)&=&\Big\{1-\frac{q^2}{4{k_F^\lambda}^2}+\frac{m^\ast v_s}
{\hbar k_F^\lambda}\frac{Q}{k_F^\lambda}
\Big(1-\lambda\sqrt{\frac{\epsilon_\alpha}{\epsilon_F+\epsilon_\alpha}}
\Big)\nonumber\\
&-&\Big(\frac{m^\ast v_s}{\hbar k_F^\lambda}\Big)^2\frac{Q^2}{q^2}
\Big(1-\lambda\sqrt{\frac{\epsilon_\alpha}{\epsilon_F+\epsilon_\alpha}}
\Big)^2\Big\}^{\frac{1}{2}}.
\end{eqnarray}
With the assumptions $q<<2k_F$ and $m^\ast v_s<<\hbar k_F$ we 
have $G(k_F,Q)\simeq 1$. Eq. (\ref{phndrag6}) can be derived easily
from Eq. (\ref{phdrag5}) with all the above mentioned approximations 
taken into account.

\section{}
In this appendix, we briefly we outline the derivation of the
energy-loss rate of hot electrons.
Using Eq. (\ref{trans}) and after doing the integration 
over $\theta$, Eq. (\ref{phn_rate}) can be re-written as 
\begin{eqnarray}\label{en_loss}
\Big[\frac{\partial N_Q}{\partial t}\Big]_{\lambda}
&=&\frac{{m^\ast}^2A}{\pi\hbar^5q}\int d\epsilon_{\bf k}
\Big(1-\sqrt{\frac{\epsilon_\alpha}{\epsilon_{\bf k}+\epsilon_\alpha}}
\Big)^2\nonumber\\
&\times&\frac{\vert C_Q\vert^2}{k^\lambda G(k^\lambda,Q)} 
N_Q(T_l)\Big[e^{(\beta_l-\beta_e)\hbar\omega_Q}-1\Big]\nonumber\\
& \times & f(\epsilon_{\bf k})\{1-f(\epsilon_{\bf k}+\hbar\omega_Q)\}.
\end{eqnarray}

At low temperature we use the same approximation made in Appendix A. 
After doing the integration over $\epsilon_{\bf k}$ in 
Eq. (\ref{en_loss}) finally Eq. (\ref{enr_relx}) becomes 
\begin{eqnarray}\label{enrel_rate}
P&=&\frac{{m^\ast}^2}{4\pi^3n_e\hbar^5}
\sum_{\lambda}\Big(1-\lambda\sqrt{\frac{\epsilon_\alpha}
{\epsilon_F+\epsilon_\alpha}}\Big)^2\int dq dq_z (\hbar\omega_Q)^2 \nonumber\\
&\times&\frac{\vert C_Q\vert^2}{k_F^\lambda G(k_F^\lambda,Q)}
N_Q(T_l)\{N_Q(T_e)+1\}\Big[e^{(\beta_l-\beta_e)\hbar\omega_Q}-1\Big].\nonumber\\
\end{eqnarray}

By taking the approximation $G(k_F,Q)\simeq 1$ into account 
at low temperature and after evaluating the integrations over 
$q$ and $q_z$ in Eq. (\ref{enrel_rate}) it
is not difficult to get Eqs. (\ref{enlsDP})-(\ref{enlsPEt}).

\end{document}